# Contractive Entanglement in Squeezed State Evolution

R. W. Rendell and A. K. Rajagopal
Naval Research Laboratory, Washington DC 20375

The position variance of single-mode contractive Yuen states can go below the standard quantum limit. For two-mode squeezed states, it is shown that the time-dependent evolution of the entanglement of formation can be contractive, going below that of the squeezed state with minimum Einstein-Podolsky-Rosen dispersions, and increasing thereafter. The rate of change of the Einstein-Podolsky-Rosen dispersions as a function of the two-mode phases control this process. Contractive entanglement is shown to be equivalent to a rotating phase space accompanied by time-dependent single-mode squeezing.

Continuous quantum variables have had a long history in the area of quantum nondemolition measurements and only more recently in quantum information processing [1, 2]. Entanglement criteria have been established for general bipartite Gaussians [3-5] and the entanglement of formation (EoF) has been derived for symmetric two-mode Gaussians [6]. A simpler sufficient condition for entanglement can be given in terms of dispersions of Einstein-Podolsky-Rosen (EPR) commuting variables [3]

$$F = Tr(\hat{\rho}\hat{F}) \equiv <\hat{F}> < 2 \qquad (1)$$

where $\hat{F} = \Delta^2(\hat{p}_1 + \hat{p}_2) + \Delta^2(\hat{q}_1 - \hat{q}_2)$. This has already been used in several experiments to demonstrate continuous variable entanglement [7]. However, entangled states are readily found which violate Eq.1. A more general inequality involving a particular class of states must be examined to raise this criterion to be necessary and sufficient [3]. Of all states with a given entanglement value, F is minimal for the two-mode squeezed states [6] (actually, for two-mode squeezed states with a particular phase, as discussed below).

Processing quantum information involves performing certain operations, such as free time evolution, controlled time evolution (e.g. quantum gates), state preparation and measurements. Dispersions have played a role in precision quantum measurements because of the interest in preparing states with uncertainties below the *standard quantum limit* (SQL). This had stimulated efforts in developing quantum non-demolition techniques and improved detectors, such as those for gravitational radiation. Yuen [8] examined free mass motion and proposed a set of *contractive states* which are capable of narrowing the position variance below the SQL bound. This was based on the observation that in the general expression for the variance:

$$\langle \Delta q^2(t) \rangle = \langle \Delta q^2(0) \rangle + t \langle \Delta q(0) \Delta p(0) + \Delta p(0) \Delta q(0) \rangle + t^2 \langle \Delta p^2(0) \rangle$$

the term linear in time can become negative due to phase effects in the q,p correlations. For example, with the initial one-mode squeezed state, $\Psi_0(q) = N \exp(-\alpha q^2/2)$ (with $\alpha = (1-\lambda)/(1+\lambda)$, $\lambda = -e^{i\varphi} \tanh(r)$, squeezing strength r and phase φ), the linear term is given by $-2\sin(\varphi)\sinh(2r)$. For $\sin(\varphi) > 0$, the position variance can be driven below the SQL for some duration of time. The variance contracts at first to a minimum and thereafter expands due to the term quadratic in time. Methods of experimentally preparing contractive states have been proposed using detuned standing light waves in a cavity [9] and stochastic cooling of optomechanical mirrors [10]. Since techniques are becoming available to produce contractive states, it is of interest to study them for more than one mode where entanglement is also present.

In this work, we explore the issues of composite states and their entanglement in the spirit of one-mode contractive Yuen states. We focus on pure symmetric two-mode Gaussian states of the form:

$$\Psi(q_1,q_2) = N \exp[-(\alpha(q_1^2 + q_2^2) + 2\gamma q_1 q_2)/2] \qquad (2)$$

where N is a normalization constant and $\alpha = \alpha_1 + i\alpha_2$, $\gamma = \gamma_1 + i\gamma_2$ are complex coefficients. Displacements are not included since they do no change the entanglement. More general aspects of asymmetric and mixed states will be published elsewhere [11]. Although the EoF of a general two-mode symmetric Gaussian can be expressed in terms of the covariance matrix transformed into standard form [6], we formulate the two-mode pure case explicitly in the coordinate representation of the wavefunction and evaluate the EoF by direct methods [11]. The EoF associated with Eq.2 can be obtained by evaluating the von Neumann entropy of the marginal one-mode states:

$$E = (\Omega + 1/2)\ln(\Omega + 1/2) - (\Omega - 1/2)\ln(\Omega - 1/2) \qquad (3)$$

where $\Omega^2 = (\alpha_1^2 + \gamma_2^2)/4(\alpha_1^2 - \gamma_1^2)$. Similarly, the expression for the EPR dispersion associated with Eq.2 is given by:

$$F = |\alpha + \gamma|^2 / (\alpha_1 + \gamma_1) + 1/(\alpha_1 - \gamma_1) \qquad (4)$$

Note that the entanglement is independent of $\alpha_2$, whereas the EPR dispersion depends on all four parameters. This is not unexpected since the EPR entanglement condition, Eq.1, is only sufficient as mentioned earlier.

In order to illustrate how the above Gaussian entanglement can exhibit contractive behavior, we prepare an initial two-mode squeezed state, which is a special form of the symmetric Gaussian, and evolve it into the more general state, Eq.2, by a nonlocal unitary evolution. This allows the entanglement between the modes to change as the state evolves. The contractive regions, which have negative position-momentum correlations, are identified and studied by way of the position variances of each mode and by the entanglement between the modes. An explicit construction is then given of local rotations and squeezings of the system phase space which allows for the state to be viewed at each time as a two-mode squeezed state but with modified phases and squeezing strengths. The EoF can be understood simply in terms of the squeezing strength of these new modes. The initial two-mode squeezed state is the two-mode squeezing operator [12], $\hat{S}_{12} = \exp(-\zeta \hat{a}_1^\dagger \hat{a}_2^\dagger + \zeta^* \hat{a}_2 \hat{a}_1)$ with $\zeta = s_0 \exp(i\varphi_0)$, applied to the vacuum. The squeezed mode annihilation operators, $\hat{S}_{12}^\dagger \hat{a}_i \hat{S}_{12}$, acting on the vacuum lead to differential equations for the wave function which is a special case of the symmetric Gaussian, Eq.2, with $\alpha=\alpha_0$, $\gamma=\gamma_0$:

$$\alpha_0 = (1 + \lambda_0^2)/(1 - \lambda_0^2), \quad \gamma_0 = -2\lambda_0/(1 - \lambda_0^2) \qquad (5)$$

and $\lambda_0 = -\tanh(s_0)\exp(i\varphi_0)$, $s_0 \geq 0$, such that $\alpha_0^2 - \gamma_0^2 = 1$. We refer to states obeying Eq.5 as the standard two-mode squeezed (STMS) state form. Here $\varphi_0$ and $s_0$ are the phase and squeezing strength of the STMS state. The phase will now be shown to play a key role in the contractive properties of the evolved state.

For the STMS state, the EoF Eq.3, takes the form:

$$E_{sq}(s_0) = \cosh(s_0)^2 \ln(\cosh(s_0)^2) - \sinh(s_0)^2 \ln(\sinh(s_0)^2) \qquad (6)$$

Note that the phase does not appear in this expression and the entanglement is controlled only by the squeezing strength. The derivation of the EoF of a general symmetric Gaussian [6] depended on the properties of a particular STMS state with $\varphi_0 = \pi$. This π-squeezed state has the important property that it has the minimum EPR dispersion F for all symmetric states with a given entanglement. Unlike the entanglement, the EPR dispersion Eq.4 generally does depend on the phase, in particular for the STMS state:

$$F_0(s_0, \varphi_0) = 2(\cosh(2s_0) + \cos(\varphi_0)\sinh(2s_0)) \qquad (7)$$

Values of $\varphi_0 \neq \pi$ lead to an increase in EPR correlations and can also lead to contractive behavior as the state evolves, as shown below.

The simplest nonlocal unitary evolution corresponds to free mass motion of the center of mass of the pair of modes: $\hat{U}(t) = \exp{-it((\hat{p}_1 + \hat{p}_2)^2/2)}$. The time-evolved wave function is obtained by applying the mode operators transformed by both $\hat{S}_{12}$ and $\hat{U}$, $\hat{U}\hat{S}_{12}^\dagger \hat{a}_i \hat{S}_{12} \hat{U}^\dagger$, to the vacuum state. This leads to differential equations for the evolved states, $\hat{U}(t)\psi_0$, of the Gaussian form Eq.2, with explicitly time-dependent coefficients:

$$\alpha(t) = \frac{1 + \lambda_0^2 + it(1-\lambda_0^2)}{1 - \lambda_0^2 + 2it(1-\lambda_0)^2}, \quad \gamma(t) = -\frac{2\lambda_0 + it(1-\lambda_0^2)}{1 - \lambda_0^2 + 2it(1-\lambda_0)^2} \qquad (8)$$

This is still a symmetric Gaussian, since $\hat{U}(t)$ is symmetric between the two modes, but is no longer in STMS form in the coordinates $q_1, q_2$ because $\alpha(t)^2 - \gamma(t)^2 \neq 1$ for t > 0. This class of states will now be shown to exhibit contractive behavior.

The time-dependences of the two-mode coordinates and momenta under $\hat{U}(t)$ are $q_i(t) = q_i(0) + (p_1(0) + p_2(0))t$, $p_i(t) = p_i(0)$ (i=1,2). This leads to expressions for the time-dependent variances. For example,

$$\langle q_1^2 \rangle_t = \langle q_2^2 \rangle_t = \cosh(2s_0)/2 + 2t\, \partial F_0/\partial \varphi_0 + t^2 F_0/2$$

This exhibits contractive behavior for $\partial F_0/\partial \varphi_0 = -\sin(\varphi_0)\sinh(2s_0) < 0$ or sin($\varphi_0$)>0, in analogy with the one-mode Yuen case. Thus contractive behavior for the case of free center of mass motion is determined by the rate of change of the EPR dispersion with phase, $\partial F_0/\partial \varphi_0 < 0$. The minimum of the contraction depends on both phase and squeezing strength and occurs at $t_m = -2(\partial F_0 / \partial \varphi_0)/F_0$.

Contractive behavior is also shown by two-mode quantities such as the EoF. Since the evolved state is no longer in STMS state form in $q_1, q_2$, the entanglement of the state cannot be directly calculated using Eq.6. For the evolved state, Eq.3 must be used instead, where $\Omega$ is now an explicit time-dependent function of the phase and squeezing strength of the initial squeezed state:

$$4\Omega(t)^2 = \cosh(2s_0)^2 + 2t\,\partial F_0/\partial \varphi_0 \cosh(2s_0) + t^2(1 + (\partial F_0/\partial \varphi_0)^2/4) \qquad (9)$$

In Fig.1, the phase dependence of the evolution of the entanglement given by Eqs.3 and 9 is displayed for $s_0 = 0.5$ and a range of phases, $\varphi_0 = \pi/4, \pi/2$ and $\pi$. At t=0, the entanglement is independent of phase as expected from Eq.6. However as the state evolves, contractive behavior occurs for $\partial F_0/\partial \varphi_0 < 0$ or $\sin(\varphi_0) > 0$ due to the linear term in $\Omega(t)$. During contraction, the states partially disentangle with increase of entanglement thereafter. The $\varphi_0 = \pi/2$ state has the maximum contraction where it becomes separable at $2t_m/(1 - \cos(\varphi_0)\tanh(2s_0)) = 2t_m$. The $\pi$-squeezed state, which has the minimum EPR dispersion, is never contractive since $\partial F_0/\partial \varphi_0 = 0$ for $\varphi_0 = \pi$. For the phases $\varphi_0 = \pi/4, \pi/2, \pi$ we find from Eq.7, the EPR dispersions $F_0 = 4.7, 3.1,$ and 0.7 respectively. For all cases of contractive entanglement, the EoF goes below that of the special $\pi$-squeezed state. The EPR dispersion, Eq.1, is invariant under the center of mass evolution so that $F(t) = F_0$ at all times. The times at which the entanglement curves cross in Fig. 1 therefore correspond to states with the same entanglement but different EPR dispersions. During the expansion phase, the entanglement continues to increase without limit, a feature which can only occur with continuous variables.

Although the EoF is given by Eqs.3 and 9, it will now be shown that we can recast it in STMS state form, Eq.6. The state in Eqs.2 and 8 is the result of the nonlocal unitary center-of-mass mass evolution but the physics of this state is found to be conceptually simpler in terms of a new set of STMS modes obtained by local transformations. The most general linear local canonical transformation $\hat{S}$ in phase space is:

$$\hat{a}'_i = \cosh(r)e^{i\theta}\hat{a}_i + \sinh(r)e^{-i\psi}\hat{a}_i^\dagger \qquad (10)$$

along with its hermitian conjugate $\hat{a}'^\dagger_i$. This is applied to the effective annihilation operator $\hat{U}\hat{S}^\dagger_{12}\hat{a}_i\hat{S}_{12}\hat{U}^\dagger$ and its hermitian conjugate. We find that at each time during the evolution, Eq.10 can transform the evolved state given by Eqs.2 and 8 into a STMS state in the new modes but with modified squeezing phase and strength. Here it is sufficient to set the phase $\psi=0$ and apply the same values of $r$ and $\theta$ to each of the two modes since the evolution is symmetric. The parameters represent local rotations by $\theta$ and local squeezings with strength $r$. Applying Eq.10, we find a new Gaussian state of the form Eq.2, but with modified coefficients:

$$\alpha'(t) = [(\Delta_+\Delta_- - \Delta_0^2)((1 + \lambda_0^2) + it(1 - \lambda_0^2)) \qquad (11a)$$
$$+ i\Delta_0\Delta_-((1 - \lambda_0^2) + 2it(1 - \lambda_0)^2) + i\Delta_0\Delta_+(1 - \lambda_0^2)]/D(t)$$

$$\gamma'(t) = -[2\lambda_0 + it(1 - \lambda_0^2)]/D(t) \qquad (11b)$$

$$D(t) = \Delta_0^2(\lambda_0^2 - 1) + 2i\Delta_0\Delta_-((1+\lambda_0^2) + it(1-\lambda_0^2)) + \Delta_-^2((1-\lambda_0^2) + 2it(1-\lambda_0)^2) \quad (11c)$$

where $\Delta_\pm = \cosh(r)\cos(\theta) \pm \sinh(r)$ and $\Delta_0 = \cosh(r)\sin(\theta)$. In order for the locally transformed state to be a STMS state, Eq.11 must obey $\alpha'(t)^2 - \gamma'(t)^2 = 1$. This time-dependent STMS condition can be met at each time and we determine the values of θ(t) and r(t) required to accomplish this. It is then described by a phase φ(t) and a squeezing strength s(t), obtained by defining:

$$\lambda'(t) \equiv -\tanh(s(t))\exp(i\varphi(t)) = (1 - \alpha'(t) - \gamma'(t))/(1 + \alpha'(t) + \gamma'(t)) \quad (12)$$

where the second equality follows from a STMS relation of the form of Eq.5.

To provide insight into φ(t) and s(t), consider the action of Eq.10 on the phase space of the center-of-mass evolved state. For a local evolution, Eq.10 would produce a modified phase, $\varphi(t) = \varphi_0 - 2(\phi_r + \theta(t))$ (where $\exp(-2i\phi_r) = (1-\lambda_r)/(1-\lambda_r^*)$, $\lambda_r = -\tanh(r)\exp(i\theta(t))$), but with s(t)=$s_0$ so that $|\lambda'(t)| = \lambda_0$. Thus the phase acquires a contribution from both the squeezing and the rotation in Eq.10 but s(t) = $s_0$ since a local evolution cannot change the entanglement. However, as shown elsewhere [11], the nonlocal aspect of $\hat{U}$ induces additional contributions to φ(t), s(t) and the entanglement. These have also been calculated numerically and are plotted in Fig.2 for the three initial phase values used in Fig.1. Both θ(t) and r(t) equal zero at t=0 where the system begins as a STMS state. However as the state begins evolving, a rapid rotation in phase space is required in order for it to remain as a STMS state. The rotation subsequently decays to zero and the squeezings increase monotonically without limit as the state continues to evolve. They exhibit a knee-like structure for the cases $\varphi_0 = \pi/4, \pi/2$ corresponding to accelerated local squeezings in the contractive region. This explicit construction of a local transformation to a STMS state is consistent with the decomposition of entangled pure Gaussians into products of squeezed states [13]. The rotations and local squeezings represented by Eq.10 can be accomplished by means of optical elements such as beam splitters, phase shifters and squeezers [14].

The time dependent phases and squeezing strengths are plotted in Fig.3 corresponding to the examples in Figs. 1 and 2 These begin at the values $\varphi_0$ and $s_0$ at t=0 corresponding to the initial STMS state. Initially the phase abruptly changes and approaches π at long times. For the cases $\sin(\varphi_0) > 0$, the phase oscillates in the interval corresponding to contractive behavior. The oscillations in φ(t) have a special character for $\varphi_0 = \pi/2$, where the EoF is smallest and vanishes at $2t_m$, and the approach to this is shown by displaying the oscillations for the cases $\varphi_0 = \pi/4, 0.4\pi$, and π/2 in Fig.3(a). Accuracy of the calculated values of the phase in the contractive region was ensured by requiring that the EPR dispersion, $F' = Tr((\hat{S}^\dagger \hat{\rho}(t)\hat{S})(\hat{S}^\dagger \hat{F}\hat{S}))$ in the transformed frame, is equal to $F_0$. The squeezing strength s(t) mimics the entanglement and also exhibits contractive behavior for these cases. Since this is a STMS state, s(t) can be used in Eq.6 to determine the entanglement and this is found to agree with Fig. 1 as it should. Thus the entanglement associated with Eqs.2 and 8 can be characterized completely by the squeezing strength s(t). Eq.11 implies that $\lambda'(t) \to 1$ as $t \to \infty$, consistent with $\varphi(t) \to \pi, s(t) \to \infty$, and $E(t) \to 2s(t)$ so that the EoF becomes directly proportional to the squeezing strength in the rotated and squeezed phase space. In terms of the original coordinates, the state asymptotically develops into the highly entangled pure phase Gaussian, $\psi_\infty = N\exp(ie^{2r}(q_1-q_2)^2/2\sin(\varphi))$ as $\varphi(t) \to \pi$, $r(t) \to \infty$, dependent only on the relative coordinates.

Quantum operations in general will transform special states (e.g. STMS) into more complicated states which have different phase dependences for the EPR dispersions and the EoF. In this paper, we illustrated this using contractive free time evolution of the position variances and the EoF of a two-mode state. The techniques used in this work may be employed in more general settings. Thus the special form of $\hat{U}(t)$ used here may in general be replaced by any operator representing quantum information processing elements. Current experimental techniques [7] have explored the demonstration of the inseparability of continuous variables [3-4] which correspond to a STMS phase, $\varphi_0=\pi$. Similar techniques could be used to explore the contractive regime corresponding to $\sin(\varphi_0)>0$ by applying single mode rotations to control the two-mode phase, $\varphi$.

**Acknowledgement:** The authors are supported in part by the Office of Naval Research.

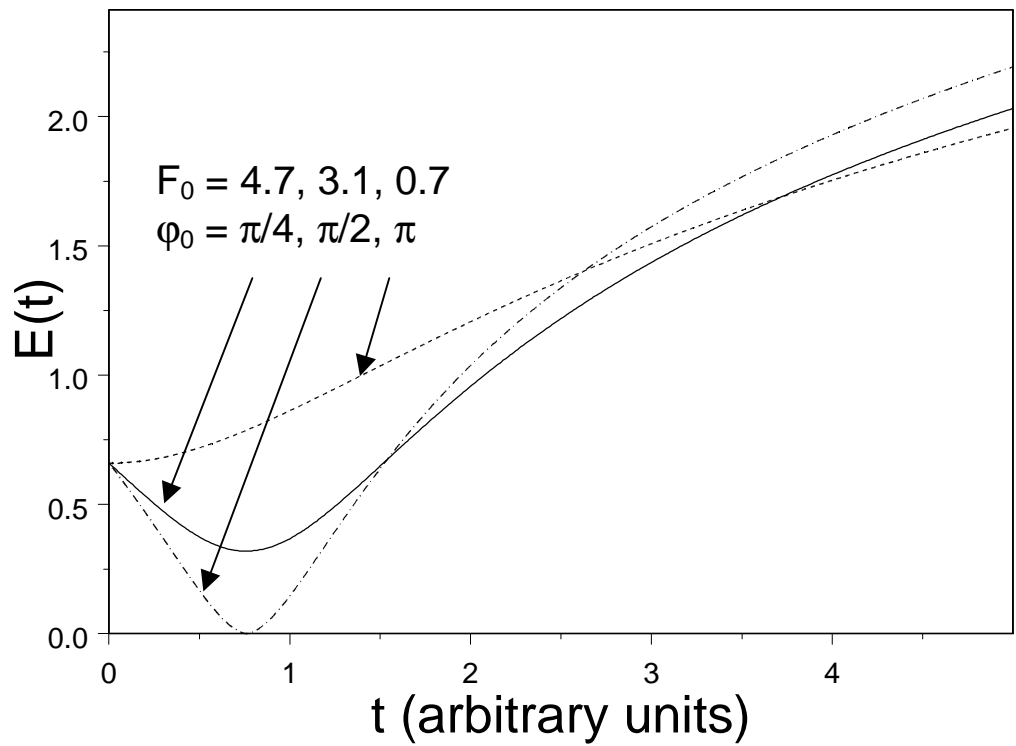

**Figure 1. Entanglement of formation for free center-of-mass evolution of initial two-mode squeezed states with $s_0 = 0.5$ and $\varphi_0 = \pi/4, \pi/2, \pi$.**

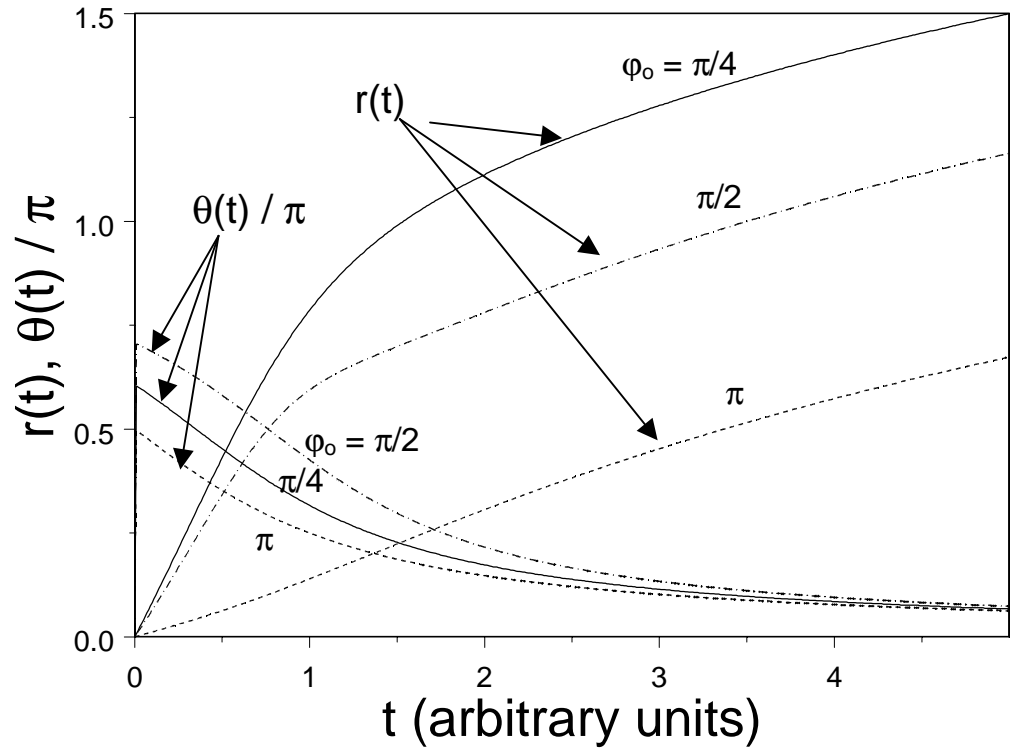

**Figure 2** The phase space rotation angle θ(t) and single-mode squeezing strength r(t) required for the free center-of-mass evolved states to be two-mode squeezed states. Parameters are as in Fig. 1.

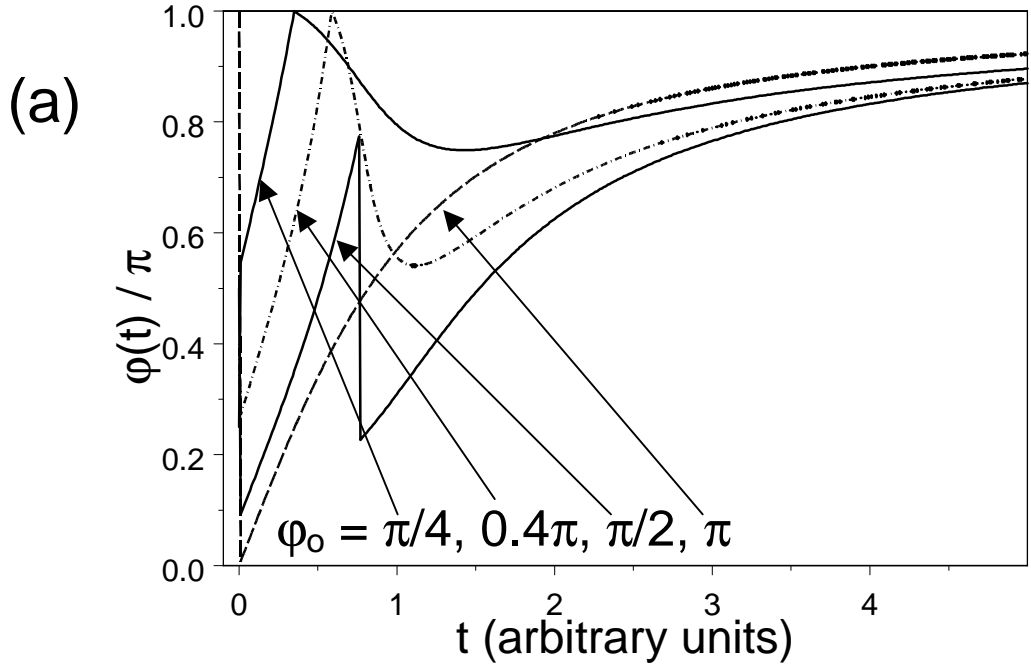

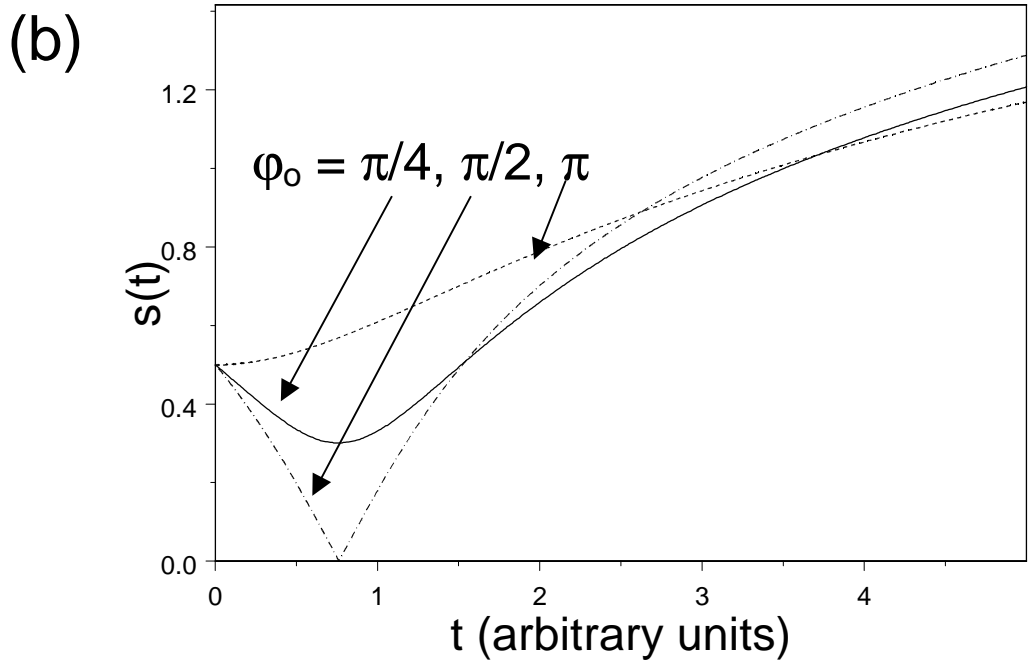

**Figure 3.** Resulting (a) phase, φ(t), and (b) squeezing strength, s(t), of the two-mode squeezed states associated with the free center of mass motion. Parameters are as in Fig.2 except for an additional value of $\varphi_0$ to show the behavior of φ(t) as $\varphi_0$ approaches π/2.